# Third nearest neighbor parameterized tight biding model for graphene nano-ribbons

Van-Truong Tran[1*], Jérôme Saint-Martin[2], Philippe Dollfus[2] and Sebastian Volz[1]

[1]EM2C, CentraleSupélec, Université Paris Saclay, CNRS, 92295 Châtenay Malabry, France
[2]C2N, Université Paris-sud, Université Paris Saclay, CNRS, 91405 Orsay, France
*vantruongtran.nanophys@gmail.com

**Abstract**

The *ab* initio band structure of 2D graphene sheet is well reproduced by the third nearest neighbor tight binding model proposed by Reich *et al* [*Phys. Rev. B* **66,** 035412]. For ribbon structures, the existing sets of tight binding parameters can successfully explain semi-conducting behavior of all armchair ribbon structures. However, they are still failing in describing accurately the slope of the bands while this feature is directly associated to the group velocity and the effective mass of electrons. In this work, both density functional theory and tight binding calculations were performed and a new set of tight binding parameters up to the third nearest neighbors including overlap terms is introduced. The results obtained with this model offer excellent agreement with the predictions of the density functional theory in most cases of ribbon structures, even in the high-energy region. Moreover, this set can induce electron-hole asymmetry as manifested in density functional theory. Relevant outcomes are also demonstrated for armchair ribbons of various widths as well as for zigzag structures, thus opening a route for multi-scale simulations.

**I. Introduction**

Density functional theory (DFT) and tight binding (TB) method are widely used to investigate and predict various properties of materials, from electronics, phononics, thermoelectrics to optics.[1–6] While the former technique does not require any empirical input parameters as it is derived directly from first principles, the latter needs several parameters such as onsite energy, hoping energy and eventually overlap terms to construct Hamiltonians.[5,7] Although the DFT usually provides relevant results compared to experimental data, it remains computationally very expensive and therefore it is only suitable for small size structures from few to few hundreds of atoms.[8] In contrast, TB models do not require self-consistent procedures for issuing the band structures, it hence consumes much less computational resources. Consequently, TB models can be implemented to examine large structures with up to millions of atoms. Additionally, TB calculations in specific cases can lead to analytical expressions which are very convenient to deepen the analysis of material's properties.[9–11] Thus DFT and TB methods have their own advantages according to the desired level of accuracy and the size of the system.

The TB parameters were usually generated by fitting TB calculations with DFT[12,13] or experimental data.[14] Although the first study of band structure of graphene was first done by

Wallace in 1947,[15] the parameterized TB model was just introduced in 1954 by Slater and Koster.[7]

To describe energy bands of 2D graphene in low energy regions, it has been shown that the $2p_z$ orbital TB model is relevant to describe important electronic properties of graphene.[16] In the case when only the nearest neighbor interaction (1NN) is considered, the nearest neighbor hoping energy to be considered is $t_1 = 2.7$ eV.[17] This simple 1NN TB model basically is accurate around the Dirac points (K points), but it is rather poor to mimic DFT or experimental data in high-energy regions.[12]

In 2002, Reich *et al* [12] have shown that the match between TB and DFT results for 2D graphene can be substantially improved by introducing a third nearest neighbor (3NN) TB model, including overlap terms. Indeed, the set of TB parameters proposed by Reich fits DFT predictions over a large range of energies. Although this set has successfully reproduced the DFT band structure, the values of the parameters seem to be mathematically fitted rather than physically as hoping and overlap of the third nearest neighbors were even larger than that of the second ones. Realizing this problem, Kundu [16] proposed to re-fit DFT data in Reich's paper and introduced a new set in which hoping and overlap terms decay as neighbor distance increases.

For the ribbon form of graphene, the simple 1NN TB model indicates that armchair graphene nanoribbons (AGNRs) of the $3p + 2$ group are semi-metallic, while AGNRs of other groups $3p + 1$ and $3p$ are actually semi-conducting.[18–20] However, DFT calculations[20] and experiments[21] have shown that all AGNRs are semiconducting. To theoretically explain the semiconducting behavior of AGNRs, Son *et al* [20] have introduced an edge deformation (ED) effect in the edges of armchair ribbons into the 1NN TB model (we note this model as 1NN + ED) and an additional term was introduced to increase the hoping between atoms at the edges up to about 12%. The added effect indeed corresponds to the underlying physics of the bandgap opening in group $3p + 2$ although it still can not replicate accurately the width of the bandgap and also the slope of energy bands. In 2008, Gunlycke and White[22] have improved the 1NN + ED model by introducing an additional term as third nearest neighbor hoping parameter $t_3$ and thus constructed a 3NN + ED model. This model can accurately reproduce the bandgap of AGNRs in most cases but still has discrepancies with the DFT data in the high-energy region of the conduction band. More important, both Son's and Gunlycke's models always present a symmetry between conduction and valence bands (electron-hole symmetry), while DFT results show that electron-hole pair is asymmetrical. When we used Reich's or Kundu's sets for ribbon calculations, the bandgap was underestimated because these sets are not optimized for ribbon structures.

Hence, it is still required to build a robust set of TB parameters that can adequately reproduce not only bandgaps but also the band shapes of nanoribbons predicted by DFT. Because the bandgap is associated directly to on/off states of devices[23] and the slope of the bands defines the group velocities and the effective masses of electrons,[24,25] the accuracy of these factors may thus affect the conclusions in a large range of transport problems.

In the present work, by implementing both DFT and TB calculations for AGNRs, we introduce a new set of parameters for 3NN TB models that presents excellent agreement with DFT results in most cases, even in high-energy regions. Furthermore, although the new set was adapted for narrow

AGNRs, we show that this set also precisely describes large armchair ribbons and zigzag structures. This modeling scheme thus opens a route for multi-scale simulations.

## II. Modeling and methodology

**Modeling:**

The sketch of both armchair and zigzag graphene nanoribbons is shown in Fig. 1, where $M$ characterizes the width of the ribbon and refers to the number of dimer lines along the width of the armchair ribbons and the number of chain lines in zigzag ribbons. The red rectangles indicate primary cells in each structure and circles indicate the ranges over which one atom interacts with its neighbors. $a_0$ denotes the nearest distance between two carbon atoms ($a_0 \sim 0.142\,\text{nm}$), the radius of the first, second and third nearest neighbor atoms are referred to as $r_1 = a_0$, $r_2 = \sqrt{3}a_0$, and $r_3 = 2a_0$, respectively.

**DFT and TB calculations:**

In this work, both DFT and TB calculations were implemented and DFT computations were implemented with the QUANTUM ESPRESSO code[26] in the framework of the plane wave basis set while TB predictions were obtained with our house code.

In DFT calculations, ribbon edges were passivated by hydrogen atoms to avoid any unexpected states generated inside the bandgap due to charge transfer induced by edge dangling bonds. In all calculations, we have also used the Perdew-Zunger (PZ) exchange-correlation functional[27] and a norm-conserving Hartwigsen-Goedeker-Hutter pseudopotential[28] within the local density approximation (LDA).[29] A kinetic energy cutoff of 90 Ry was chosen to safely converge total energies. To mesh the Brillouin zone for integrals, a Monkhorst-Pack $40\times1\times1$ was used.[30] All structure were relaxed until the force on each atom was less than 0.001 Ry.au$^{-1}$.

Regarding TB calculations, we start with general Bloch wave functions in periodic structures [7]:

$$\left|\Psi_{\vec{k}}\left(\vec{r}\right)\right\rangle = \frac{1}{\sqrt{N}} \sum_{\beta=-\infty}^{+\infty} \sum_{j=1}^{P} e^{i\vec{k}\cdot\vec{R}_\beta} \cdot \left(c_j \left|\varphi_j\left(\vec{r}-\vec{R}_\beta\right)\right\rangle\right), \qquad (1)$$

where $N$ is the number of unit cells in the crystal, $P$ is number of atoms in a unit cell and $\left|\varphi_j\left(\vec{r}-\vec{R}_\beta\right)\right\rangle$ is the $2p_z$ orbital at atom $j$-th in the unit cell $\beta-\text{th}$.

The time-independent Schrödinger equation is commonly written as

$$H\left|\Psi_{\vec{k}}\left(\vec{r}\right)\right\rangle = E\left|\Psi_{\vec{k}}\left(\vec{r}\right)\right\rangle \qquad (2)$$

Substituting (1) into (2) and multiplying both sides by $\langle \varphi_i(\vec{r}-\vec{R}_\alpha)| e^{-i\vec{k}.\vec{R}_\alpha}$ yields

$$\sum_{\beta=-\infty}^{+\infty} e^{i\vec{k}.(\vec{R}_\beta-\vec{R}_\alpha)} \sum_{j=1}^{P} H_{i\alpha,j\beta}.c_j = E \sum_{\beta=-\infty}^{+\infty} e^{i\vec{k}.(\vec{R}_\beta-\vec{R}_\alpha)} \sum_{j=1}^{P} S_{i\alpha,j\beta}.c_j, \quad (3)$$

where $H_{i\alpha,j\beta} = \langle \varphi_i(\vec{r}-\vec{R}_\alpha)|H|\varphi_j(\vec{r}-\vec{R}_\beta)\rangle = -t_{i\alpha,j\beta}$ is the Hamiltonian element directly associated to the hoping coupling between atom $i$ of the $\alpha-$th unit cell and atom $j$ of the $\beta-$th unit cell and $S_{i\alpha,j\beta} = \langle \varphi_i(\vec{r}-\vec{R}_\alpha)|\varphi_j(\vec{r}-\vec{R}_\beta)\rangle = s_{i\alpha,j\beta}$ is the overlap of the two wave functions. In a general 3NN TB parameterized model, $\{t_{i\alpha,j\beta}, s_{i\alpha,j\beta}\}$ will be fitted to $\{t_1, s_1\}, \{t_2, s_2\}$ or $\{t_3, s_3\}$ depending on the distance between atoms $i$ and $j$. In the case where $\alpha \equiv \beta$ and $i \equiv j$, $H_{i\alpha,i\alpha} = \langle \varphi_i(\vec{r}-\vec{R}_\alpha)|H|\varphi_i(\vec{r}-\vec{R}_\alpha)\rangle = E_{2p}$ is the $2p_z$ onsite energy of a carbon atom, and obviously $S_{i\alpha,i\alpha} = \langle \varphi_i(\vec{r}-\vec{R}_\alpha)|\varphi_i(\vec{r}-\vec{R}_\alpha)\rangle = 1$.

Combining the $P$ equations constructed from equation (3) (as $i = 1:P$), a matrix equation can be formed as

$$\left( H_{\alpha\alpha} + \sum_{\beta\neq\alpha} H_{\alpha\beta}.e^{i\vec{k}.(\vec{R}_\beta-\vec{R}_\alpha)} \right)\phi_0 = E.\left( S_{\alpha\alpha} + \sum_{\beta\neq\alpha} S_{\alpha\beta}.e^{i\vec{k}.(\vec{R}_\beta-\vec{R}_\alpha)} \right)\phi_0, \quad (4)$$

where $H_{\alpha\beta} = \{H_{i\alpha,j\beta}\}, S_{\alpha\beta} = \{S_{i\alpha,j\beta}\}$ are the matrices containing all interactions of atoms between the two $\alpha-$th and $\beta-$th cells and $\phi_0 = (c_1 \quad c_2 \quad ... \quad c_P)^T$.

Setting $H = H_{\alpha\alpha} + \sum_{\beta\neq\alpha} H_{\alpha\beta}.e^{i\vec{k}.(\vec{R}_\beta-\vec{R}_\alpha)}$, $S = S_{\alpha\alpha} + \sum_{\beta\neq\alpha} S_{\alpha\beta}.e^{i\vec{k}.(\vec{R}_\beta-\vec{R}_\alpha)}$, leads to the Eigenvalue problem which provides the band structure:

$$(S^{-1}H)\phi_0 = E\phi_0 \quad (5)$$

**Density of States calculations:**

To obtain the density of states (DOSs) in the frame of the TB method, we used the Gaussian smearing of the delta function, i.e. : [31]

$$D(E) = \sum_n \sum_{\vec{k} \in BZ} \delta\left(E - E_n(\vec{k})\right) = \sum_n \sum_{\vec{k} \in BZ} \frac{1}{\eta\sqrt{\pi}} \cdot e^{-\frac{\left(E - E_n(\vec{k})\right)^2}{\eta^2}}, \tag{6}$$

where $n$ refers to the band index and $\eta$ to a small positive number.

To reach a high resolution for the DOSs, we used in both DFT and TB calculations a $1000 \times 1 \times 1$ $k$-mesh grid for implements of DOS calculations.

## III. Results and discussion

### 1. Assessing existing sets of TB parameters for ribbon calculations

We first examine the merit of existing sets of TB parameters for ribbon structures. In Fig. 2 we display the band structure and the DOS calculated by DFT and different TB models. Two different groups of sets of TB parameters were distinguished: in Fig. 2(a), we employed three different sets of TB parameters which have been fitted for a 2D graphene sheet, while in Fig. 2(b), two sets of TB parameters proposed by Son and Gunlycke for AGNRs were used. The parameter values for each set are reported in Table 1.

In Fig. 2(a), the simple 1NN TB model with taking into account only one nearest hoping $t_1$ induces energy bands (solid pink lines) with a zero bandgap in an armchair ribbon of width $M = 5$. This gapless characteristic has been also predicted for all ribbons of group $M = 3p + 2$ as reported in refs.[18,19] In contrast, results from the 3NN TB models by both Reich's [12] and Kundu's [16] sets indicate that the structure is a semiconductor, which is thus in agreement with conclusions of DFT data.[20] However, both sets lead to smaller gaps compared to that deduced from the DFT, which is reflected clearly in the inset of the panel of the DOSs.

Although Kundu's set was shown to agree with the DFT band structure of 2D graphene, and particularly to be more physically grounded than the set of Reich, with hoping and overlap parameters decaying for longer neighbor distances, observing the energy bands and the DOS shows that the set of Reich conforms to DFT results better than the one of Kundu in the armchair ribbon structure with $M = 5$. The weakness of the Kundu set when applied to ribbons may be related to the fact that rule of decay versus distance of hoping and overlap parameters are different for atoms inside the ribbons and for the ones near the edges. We hence should only consider the average effect of this rule.

We also observe a discrepancy in the vicinity of the bandgap when applying these two sets to ribbons of different widths (not shown). In fact, these sets cannot accurately reproduce the bandgap and also the shape of the bands in ribbons structure because they have been fitted for 2D graphene sheet only. As a consequence, they do not include properly the ribbon's finite size effects which are strongly influencing the bandgap.

It is also worth noting that the highest peaks appearing in the DOS are due to flat bands resulting from the simple 1NN TB model.

To overcome the failure of the 1NN TB model in reproducing the bandgap of AGNRs, Son[20] has proposed a deformation effect in the edges of the ribbon and added a term in the model to describe this effect. In this corrected model, the nearest hoping coupling at the edge increases by 12% compared to the one in bulk. As a result, this effect induces a finite bandgap in ribbons of group $M = 3p + 2$ as confirmed in the left and the right panels of Fig. 2(b) (yellow lines). Essentially, this correction turns a semi-metallic ribbon into a conducting one and agrees well with the DFT and experimental results. However, the accuracy of the bandgap and the band shape still needs to be improved. As can be clearly observed in the inset of the Fig. 2(b) in the right, the DOS arising from the 1NN TB model shows a mismatch with DFT results around the bottom of the first conduction band ($E_{1c}$) and the top of the first valence band ($E_{1v}$).

In the model of Gunlycke,[22] the problem of the bandgap was greatly improved as the authors introduced an additional term corresponding to the third nearest neighbor hoping energy. As it will be shown later, this added term is a pertinent choice by solely adding ED effects and correct TB method predictions to be closer to the DFT ones. From the DOS lines, we note that the 3NN TB model using Gunlycke's set gives a value of bandgap in very good agreement with that of DFT in this armchair structure. However, the imparity in conduction bands remains noticeable even in the low energy range from 0.4 eV to 1eV.

More importantly, both TB models for ribbons yield a symmetry between conduction and valence bands via the line aligned to the middle of the bandgap. However, DFT clearly renders an electron-hole asymmetry, which cannot be reproduced by these existing sets of TB parameters.

A better set is therefore needed not only to reproduce the bandgap but also the shape of bands playing a significant role in high-energy transport problems as it is directly associated to the group velocity and the effective mass of carriers.

## 2. Impact of TB parameters on the energy bands

To understand which term needs to be introduced to make predictions of TB method closer to the DFT ones, we first examine the difference between, on one hand, conduction bands and valence bands resulting from DFT and, on the other hand, the outcomes of the simple 1NN TB technique (only $t_1$ involved). Due to the multiplicity of the bands, we simplify the calculations by only considering the difference in the lowest conduction band ($\Delta E_{1v}$) and the highest valence band ($\Delta E_{1c}$) as they are the bands that most contribute to transport.

As it can be seen in Fig. 3(a), the first conduction band of DFT (red line with open circles) is higher than the one of the 1NN TB model in a short range of energy in the vicinity of the Gamma point. But it becomes lower than its counterpart of the 1NN TB model for higher values of k-points. An inverse result is observed for the first valence band. These considerations can also been deduced from Fig. 2(a).

By adding different terms into the 1NN TB model, we could understand which parameter is the relevant term to be added in order to mimic the outcomes of DFT calculation.

We consequently introduced six new sets of parameters with set 1 being constructed by introducing an edge relaxation $\Delta t_1 = 0.12 t_1$ which is thus actually the Son's model. Other sets are corresponding to the addition of $t_2, t_3, s_1, s_2$ or $s_3$, respectively.

We have calculated again $\Delta E_{1v}$ and $\Delta E_{1c}$ to compare the results induced by these new sets of parameters to the simple 1NN TB model. The results are presented in Fig. 3(b).

At the first glance, visible gaps are observed when introducing either edge relaxation $\Delta t_1$ (Fig. 3(b1)) or hoping of the third nearest neighbors $t_3$ (Fig. 3(b3)). Adding another term such as $s_1$ or $s_3$ brings almost not change in the bandgap as $E_{1v}$ and $E_{1c}$ are preserved at the Gamma point but strongly alters the bands at high $k$-point values. Although the adding of $t_2$ leads to a shift of all bands, the equality of $\Delta E_{1v}$ and $\Delta E_{1c}$ at $k = 0$ indicates that the bandgap remains the same (equal to 0) as in the 1NN model. The introduction of $t_2$ is actually equivalent to adding a varying spacing in potential energy on each lattice site to shift and module the band structure.

Adding $s_2$ also retains the bandgap as $\Delta E_{1c} = \Delta E_{1v} = 0$ at $k = 0$. However, for other structures of different widths such as $M = 6$ or $M = 7$, we observe that adding $s_2$ leads to an opening of the energy gap (not shown).

It has been shown in ref.[25] that the effects of adding either the edge relaxation term $\Delta t_1$ or the hopping energy $t_3$ are equivalent regarding the opening of a bandgap. However they are actually equivalent only in the vicinity of $k = 0$. Indeed, they both show a gap opening, but away from the Gamma point, adding $\Delta t_1$ yields a shift up (down) of the conduction (valence) bands. Whereas, for high $k$-point values, adding $t_3$ shifts down (up) the conduction (valence) bands as $\Delta E_{1v} < 0$ ( $\Delta E_{1c} > 0$). Adding $t_3$ finally appears more relevant to match the results of DFT shown in Fig. 3(b) than by introducing $\Delta t_1$.

It is also very important to note that Figs. 3(b1), 3(b3), 3(b5) depicting the effect of introducing the ED term $\Delta t_1$, $t_3$, and $s_2$, respectively, reveals that these parameters do not break the symmetrical property between conduction and valence bands, as in the simple 1NN TB model. In contrast, the introduction of $t_2$, $s_1$ and $s_3$ as presented in Figs. 3(b2), 3(b4), and 3(b6), respectively, can induce an asymmetrical behavior of conduction and valence bands.

As it can be seen from the DFT energy bands in Fig. 2, electron-hole symmetry should not be expected. In consequence, adding $t_3$ or/and $\Delta t_1$ is not enough to precisely mimic the band shapes predicted by DFT. This outcomes justifies why both Son's and Gunlycke's models are failing in reproducing the bands of DFT although they successfully explain semiconducting behaviors of the ribbons of group $M = 3p + 2$. Accordingly, it is mandatory to also introduce overlap terms in order to optimize the accuracy of the TB technique against DFT calculation.

**3. The new 3NN TB model for graphene ribbons**

Following the idea that adding $t_3$ is more relevant than introducing $\Delta t_1$, we use a general 3NN TB model without introducing $\Delta t_1$ to aim at the best fit between TB calculations and DFT data.

We start with a reasonable guess set of 3NN TB parameters, $E_{2p} = -0.187$ eV being qualitatively accurate to reproduce the DFT bandgap of the structure $M = 5$. Then we vary the other six parameters including $t_1, t_2, t_3, s_1, s_2$ and $s_3$, around the guess values, i.e, we vary eleven values around the guess value of each parameter and construct in total $11^6$ sets to be scanned. We then calculated the energy error of each bands provided by the TB model with each set compared to DFT data. Due to the large number of bands, we considered the error for only the first conduction and valence bands, which are the most relevant to transport properties. The set providing the lowest error was then selected and shown in Table 1.

We employed the parameters of the selected set and plotted the corresponding energy bands and compared them to the DFT ones. The comparison is displayed in Fig. 4 where the DOSs were also reported for further comparison. As clearly manifested from the band structure and the DOS panels, TB calculations (dashed red lines) obtained from the set proposed in this work show an excellent agreement with DFT results for energies ranging from -3 eV to 3 eV. Around the region of the bandgap, the inset of the DOSs indicates the high accuracy of TB data as both the bandgap and the DOSs are identical to the DFT results. This new set of parameters seems to be well optimized compared to the existing TB ones proposed by Reich[12] and Kundu[16] fitted for 2D graphene and Son[20], Gunlycke[22] fitted for ribbons structures.

To further validate this new set of parameters, we compare results of the band structure obtained from DFT, 3NN TB with Gunlycke's set and 3NN TB model with the set of this work. First, results for other groups of ribbons, i.e. $M = 6$ (group $3p$) and $M = 7$ (group $3p + 1$), are shown to support the robustness of the new set.

In Figs. 5(a) and 5(b), those results are obviously better than the ones of the Gunlycke's set and, overall, they fit very accurately with DFT outcomes although both Gunlycke's set and the new set have bandgaps slightly smaller than the one of DFT in the case of $M = 6$. Though the new set was constructed from a fit in a narrow structure with $M = 5$, Figs. 5(c) and 5(d) displaying the results obtained for $M = 11$ and $M = 19$, respectively, show that this set still has strong relevance for larger ribbons. Band structures resulting from our model (red lines) still fit very precisely DFT data (black circles) even in the high energy regions. In the large ribbons, the results of the Gunlycke set (blue) are good in the low energy regions around the gap and in a large part of the valence band. However, it still exhibits a substantial inconsistency with DFT results in the high-energy regions of the conduction bands.

Additionally, we have also checked the set for zigzag structures. The band structure for a zigzag ribbon of width $M = 11$ chain lines along the width is displayed in Fig. 6, including both DFT and TB results. The new TB set shows a very good agreement with DFT, particularly in the first conduction and valence bands and at the energy points at the boundary of the Brillouin zone. Gunlycke's set also matches well with DFT predictions for the first valence bands but obviously poorly fits in the high-energy region of the conduction bands. These results reinforce the relevance of the new set of TB parameters proposed in the present work.

## IV. Conclusion

In conclusion, we have analyzed the relevance of commonly used sets of TB parameters for graphene structures. We have shown that although 3NN TB parameters such as Reich's or Kundu's ones have been demonstrated to accurately fit DFT bands of 2D graphene, they are not efficient in reproducing DFT predictions for ribbon structures. Other parameters of Son and Gunlycke were optimized for armchair ribbons but these sets of parameters can not mimic the electron-hole asymmetry of the energy bands. By implementing both DFT and TB calculations and then introducing a fit, we have shown the superior accuracy of a new set of TB parameters including up to 3NN plus overlap terms for graphene ribbons. The new set has been demonstrated to be in excellent agreement with DFT results in most cases, even in high energy regions and for large ribbons. Although the set has been fitted for armchair ribbons, it has been shown to also correctly describe zigzag structures.

## Acknowledgments

This work was supported by the TRANSFLEXTEG EU project.

Table 1: Different sets of tight binding parameters for 2D and ribbon graphene structures.

| Fit for structure | Set | $E_{2p}$ (eV) | $t_1$ (eV) | $t_2$ (eV) | $t_3$ (eV) | $s_1$ | $s_2$ | $s_3$ | $\Delta t_1$ (eV) |
|---|---|---|---|---|---|---|---|---|---|
| 2D Graphene | 1NN | 0 | 2.7 | | | | | | |
| | 3NN (Reich2002) | -0.28 | 2.97 | 0.073 | 0.33 | 0.073 | 0.018 | 0.026 | |
| | 3NN (Kundu2011) | -0.45 | 2.78 | 0.15 | 0.095 | 0.117 | 0.004 | 0.002 | |
| Graphene ribbons | 1NN+ED (Son2006) | 0 | 2.7 | | | | | | 0.12*$t_1$ |
| | 3NN+ED (Gunlycke2008) | 0 | 3.2 | | 0.3 | | | | 0.0625*$t_1$ |
| | 3NN (this work) | -0.187 | 2.756 | 0.071 | 0.38 | 0.093 | 0.079 | 0.070 | |

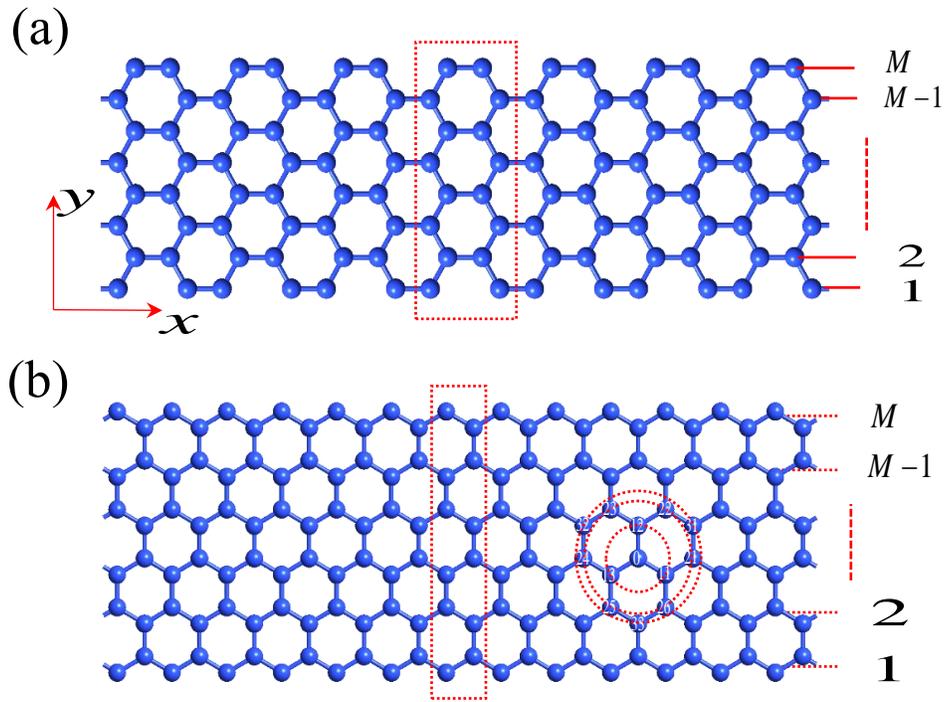

Fig 1. Two typical ribbon structures with (a) armchair edges and (b) zigzag edges. The width of the ribbons is characterized by the parameter *M* referring to the number of dimer lines in armchair ribbons and to the number of chain lines in zigzag ribbons.

(a)

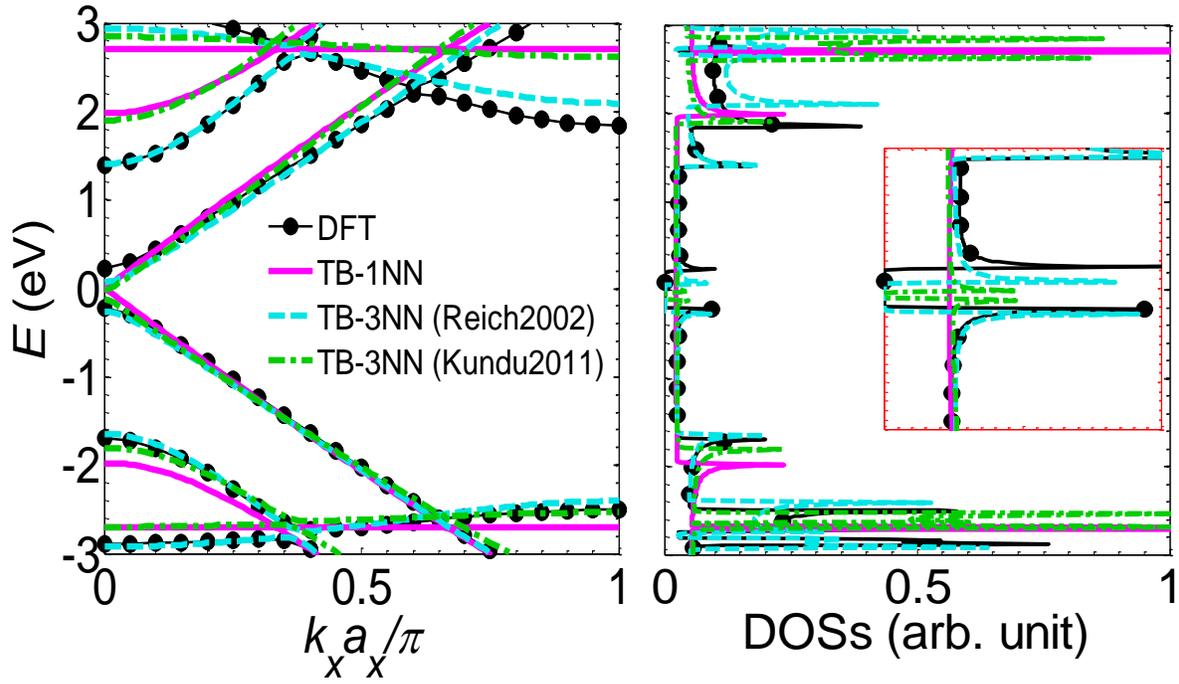

(b)

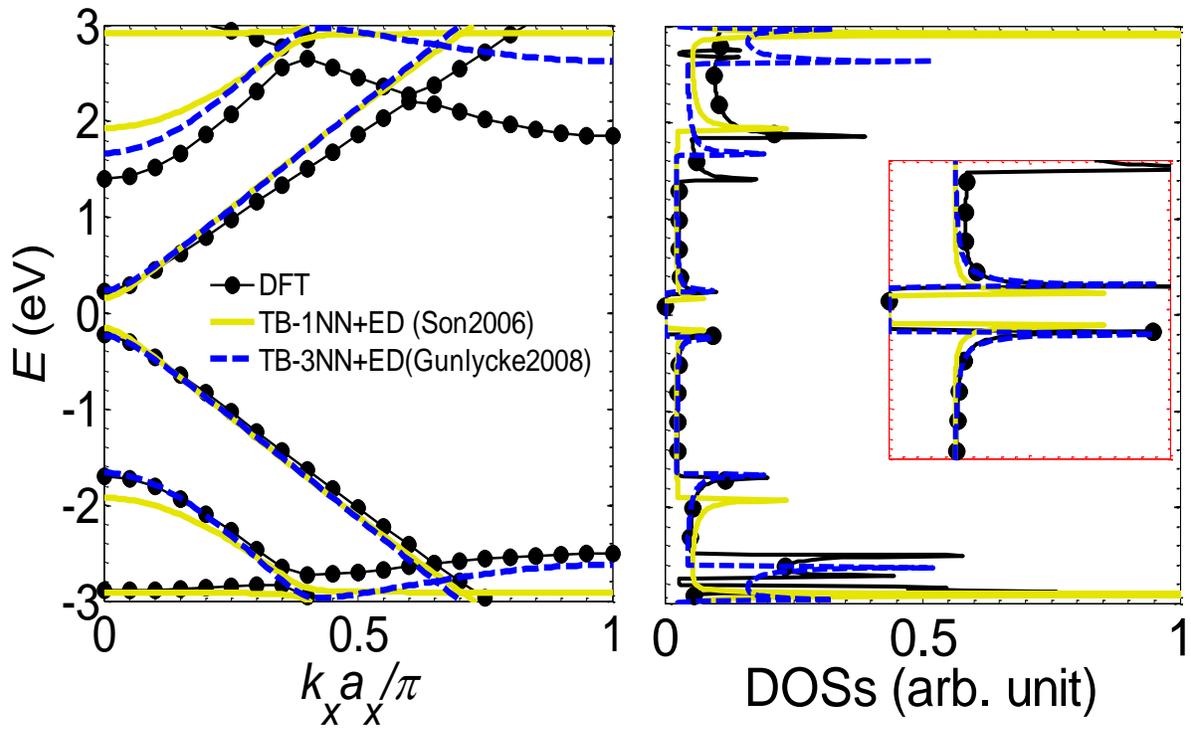

Fig 2. Comparison of energy bands and the density of states (DOSs) of an armchair ribbon of width $M = 5$ for data calculated by DFT and TB method. (a) The sets of TB parameters fitted for 2D graphene were used. (b) The sets of TB parameters fitted for graphene ribbons were used.

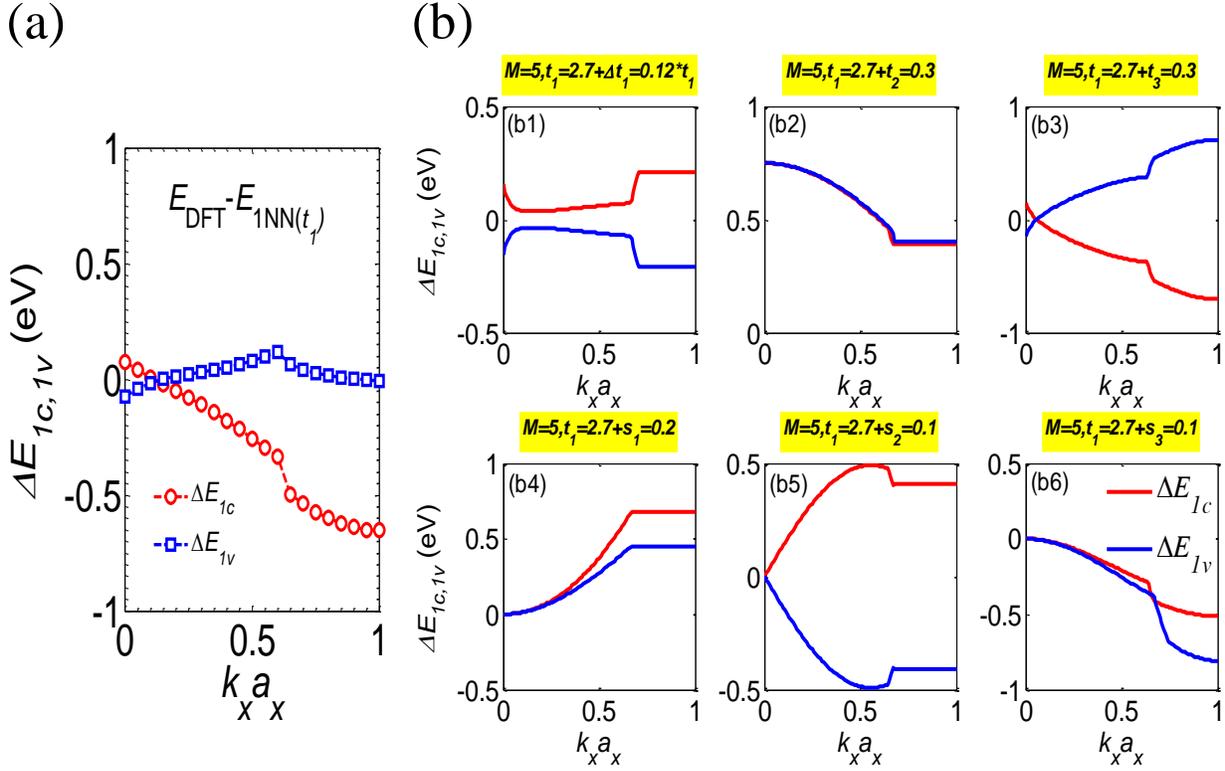

Fig 3. (a) The difference of energy in the first conduction ($\Delta E_{1c}$) and first valence ($\Delta E_{1v}$) bands of DFT and simple 1NN TB calculations. (b) $\Delta E_{1c}$ and $\Delta E_{1v}$ were calculated to compare results of the new set of parameters and the simple 1NN TB model. All calculations are performed for the armchair ribbon $M = 5$.

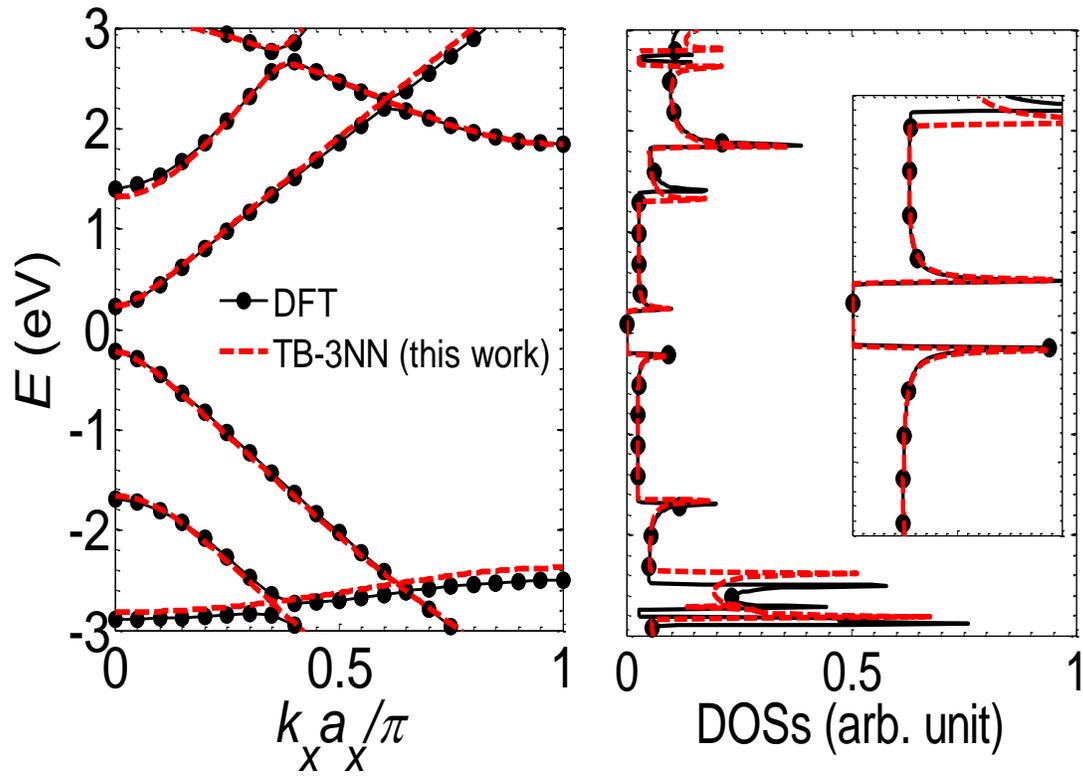

Fig 4. (a) Energy bands and (b) the density of states (DOSs) calculated for the armchair ribbon $M$ = 5 by DFT and 3NN TB model with the new set of parameters proposed in this work.

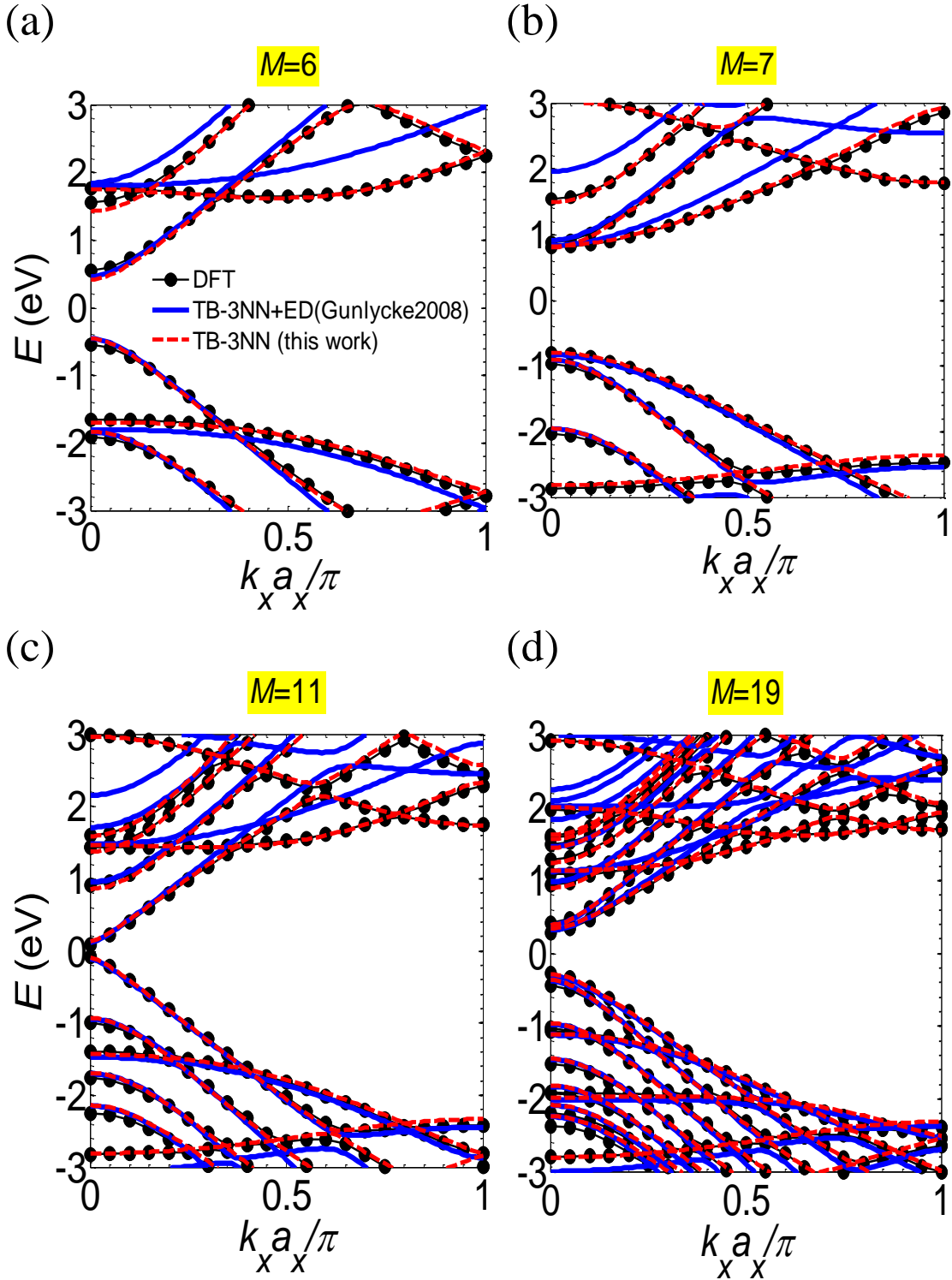

Fig 5. Comparison of energy bands calculated by DFT (black filled circle), the 3NN TB model using the set of Gunlycke (blue) and the 3NN TB model using the set proposed in this work (red) for different ribbons of width: (a) $M = 6$, (b) $M = 7$, (a) $M = 11$, and (d) $M = 19$.

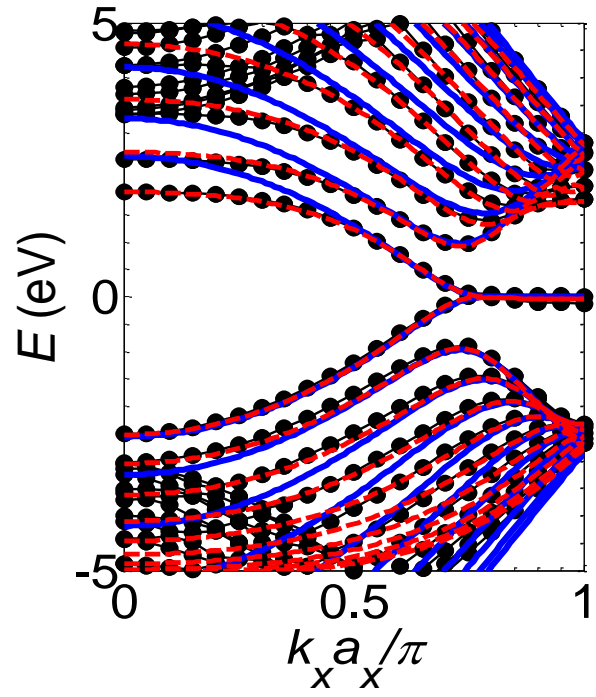

Fig 6. Comparison of energy bands for a zigzag graphene ribbons of width $M = 11$ with DFT (black with filled circle ), the 3NN TB model using the set of Gunlycke (blue) and the 3NN TB model using the set proposed by this work (red).